\documentclass[aps,twocolumn,prl,superscriptaddress,footinbib]{revtex4-1}
\usepackage{amsmath,amssymb,bm}
\usepackage{graphicx}
\usepackage{epstopdf}
\usepackage{latexsym}
\usepackage[normalem]{ulem} 
\usepackage[usenames,dvipsnames]{color}
\usepackage{hyperref}
\usepackage{natbib}
\usepackage{braket}
\begin{document}

\newcommand{\Rev }[1]{{\color{blue}{#1}\normalcolor}} 
\newcommand{\Com}[1]{{\color{red}{#1}\normalcolor}} 

\title{A cavity-QED protocol for precise field sensing in the optical domain}
\date{\today}

\author{Robert J. Lewis-Swan}
\affiliation{JILA, NIST, Department of Physics, University of Colorado,  Boulder, CO 80309, USA}
\affiliation{Center for Theory of Quantum Matter, University of Colorado, Boulder, CO 80309, USA}
\author{Diego Barberena}
\affiliation{JILA, NIST, Department of Physics, University of Colorado,  Boulder, CO 80309, USA}
\affiliation{Center for Theory of Quantum Matter, University of Colorado, Boulder, CO 80309, USA}
\author{Juan A. Muniz}
\affiliation{JILA, NIST, Department of Physics, University of Colorado,  Boulder, CO 80309, USA}
\author{Julia R.~K. Cline}
\affiliation{JILA, NIST, Department of Physics, University of Colorado,  Boulder, CO 80309, USA}
\author{Dylan Young}
\affiliation{JILA, NIST, Department of Physics, University of Colorado,  Boulder, CO 80309, USA}
\author{James K. Thompson}
\affiliation{JILA, NIST, Department of Physics, University of Colorado,  Boulder, CO 80309, USA}
\author{Ana Maria Rey}
\affiliation{JILA, NIST, Department of Physics, University of Colorado,  Boulder, CO 80309, USA}
\affiliation{Center for Theory of Quantum Matter, University of Colorado, Boulder, CO 80309, USA}


\begin{abstract}

In the context of quantum metrology, optical cavity-QED platforms have  primarily been focused on the generation of entangled atomic spin states useful for next-generation frequency and time standards. Here, we report a complementary application: The use of  optical cavities to generate non-classical states of light for electric field sensing below the standard quantum limit. We show that cooperative atom-light interactions in the strong collective coupling regime can be used to engineer generalized atom-light cat states which enable quantum enhanced sensing of small displacements of the cavity field even in the presence of photon loss. We demonstrate that metrological gains of $10-20$ dB below the standard quantum limit are within reach for current cavity-QED systems operating with long-lived alkaline-earth atoms. 

\end{abstract}

\maketitle  

\noindent{\textit{Introduction:}} The advent of quantum technologies promises to bring with it significant advances in quantum  computation, simulation and metrology among others, by utilizing resources such as entanglement and  many-body coherence  to outperform classical analogs. For example, entangled quantum states facilitate precise sensing of weak perturbations, beyond the so-called standard quantum limit (SQL), which bounds the performance of classical sensors. 

The SQL is defined as the precision to which a parameter can be estimated using only quasi-classical states such as bosonic coherent states \cite{Braunstein1994,Jaekel_1990}. The isotropic rms width of the coherent state phase-space distribution \cite{walls_quantum_2008} sets a lower bound on how well a small perturbation to a system can be estimated. For example a small displacement $\beta$ of the coherent state $\vert \alpha \rangle$ can be inferred to $(\delta\beta)^2_{\mathrm{SQL}} = 1/4$ \cite{Caves_1980}. Importantly, the sensitivity is independent of the mean number of particles $\bar{n} = \vert\alpha\vert^2$ in the coherent state, implying there is no simple improvement  by using classical states with higher occupation. However, by introducing correlations and entanglement, the quantum projection noise of a bosonic state can be manipulated to achieve precision beyond the SQL to the Heisenberg limit \cite{Zurek2001,Pezze_2018,Giovannetti_2006,Yurke_1986,Holland_1993}, which scales as $(\delta\beta)^2 \sim 1/\bar{n}$ for small displacements. Experiments with Rydberg atoms \cite{Penasa_2016,Facon2016} have demonstrated sub-SQL precision in the microwave domain and relevant progress has also been made with superconducting qubits \cite{Vlastakis2013} and trapped ions \cite{Didi_2018, Burd2019}. Potential applications include detection of single electrons \cite{Devoret2000,Atature_2011} and photons \cite{Hempel_2013}, searches for dark matter \cite{Malnou2019}, and quantum information processing \cite{Ladd2010}.

To reach sub-SQL capabilities using matter-light interactions in single qubit systems requires experiments to operate in the  strong coupling regime of cavity-QED,  achieved when the matter-light coupling rate, $2g$, is larger than the decay rates of the qubit, $\gamma$, and the cavity, $\kappa$  \cite{Schuster_2007,Girvin_2014,Hacker_2019}. In state-of-the-art optical cavities this limit can be hard to attain. However, here we demonstrate that even when $g\ll \kappa$, the interrogation of a collective ensemble coupled to a single cavity mode can be used for the preparation of  generalized atom-light cat-states and quantum-enhanced sensing in the optical regime. Although such states are hard to detect and manipulate we discuss an interferometric protocol based on time-reversal of the dynamics, which allows for nearly optimal metrological performance using accessible observables such as atomic inversion.

Our observations are relevant and directly applicable to current state-of-the-art optical cavities coupled to optical transitions in alkaline earth atoms, such as the narrow ${}^1$S$_0$-${}^3$P$_1$ optical transition in $^{88}$Sr. We predict in this system it will be possible to reach $10-20$~dB below the SQL operating with $\sim 10^5$ atoms. 
Our protocol thus opens a path for sub-SQL sensing of electromagnetic fields in the optical domain, and could help to circumvent shot-noise limitations in interferometry \cite{Kimble1987,Grangier1987,Lam2002,Taylor2013,Weaver_2018,AdvancedLigo2016,Dooley_2016}, as well as in other frequency regimes since it is applicable to a broad range of platforms featuring similar types of atom-light couplings \cite{Bollinger_2018, Delglise_2008,Wallraff_2009,Viennot_2018,Fink2009,Kolkowitz_2012,Aspelmeyer_2014,Gilmore2017,Burd2019}.

\begin{figure}[!t]
     \includegraphics[width=8cm]{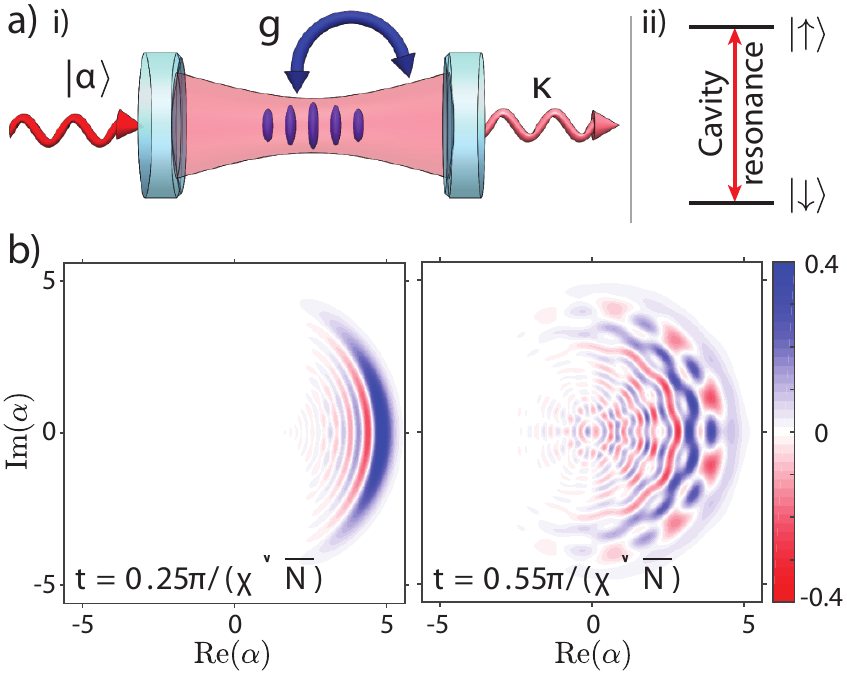}
    \caption{(a) Proposed cavity-QED setup. $N$ atoms are trapped in a standing-wave optical lattice. The optical transition of each atom forms a pseudospin-$1/2$ which is coupled to the field of the resonant optical cavity according to Eq.~(\ref{IIHamiltonian}). Photon leakage from the cavity at rate $\kappa$ is the dominant decoherence mechanism. (b) Wigner function of equivalent cavity field generated by dispersive interaction, $\vert \phi_{\mathrm{cav}}\rangle \propto \sum_{m = -N/2}^{N/2} c_m \vert \alpha e^{-i\omega_m t} \rangle$ where $c_m$ are adopted from a coherent spin-state [see definition of Eq.~(\ref{genericCat})] and $\omega_m = \chi m$. Increasingly fine structure emerges in the phase-space distribution as time proceeds. We choose $N = 10$ and $\alpha = 4$ for illustration. }
    \label{fig:CavityAndWigner}
\end{figure}

\noindent{\textit{Model:}} We seek to realize a dispersive atom-light coupling between a single mode of an optical cavity and an ensemble of atoms each encoding a spin-$1/2$ degree of freedom in an optical transition [Fig.~\ref{fig:CavityAndWigner}(a)], described by the Hamiltonian:
\begin{equation}\label{IIHamiltonian}
\hat{H}=\chi \hat{a}^{\dagger} \hat{a} \hat{S}_x. 
\end{equation}
Here, $\hat{a}$ ($\hat{a}^{\dagger}$) are destruction (creation) operators for the cavity mode, $\hat{S}_{x,y,z}=\sum_{j=1}^N\frac{\hat{\sigma}_{x,y,z}^j}{2}$ are collective spin operators with $\hat{\sigma}_{x,y,z}^j$ Pauli operators acting on atom $j$ and we set $\hbar = 1$ throughout this manuscript. This interaction is a generalization of that commonly engineered in microwave cavity and circuit-QED platforms \cite{Bertet_2002,Schuster_2007,Blais_2004,Girvin_2014} and similar to that engineered in optomechanics \cite{Aspelmeyer_2014}. Whilst decoherence will play an intrinsic role in any practical realization, we will first focus on the non-classical states which can ideally be generated via the coherent dynamics described by Eq.~(\ref{IIHamiltonian}). Later in the manuscript we will discuss how this effective Hamiltonian can be engineered in an optical cavity, by injecting a large coherent displacement into the cavity tuned to be resonant with the atomic transition [Fig.~\ref{fig:CavityAndWigner}(a)].

The dispersive atom-light interaction can be interpreted as an $\hat{S}_x$-dependent rotation of the cavity field, or alternately a precession of the spin degree of freedom at a rate controlled by the cavity occupation. The former understanding motivates its use in the  generation of entangled cat states composed of superpositions of bosonic coherent states, as has previously been demonstrated in the case of single Rydberg atoms in a microwave cavity \cite{Penasa_2016}. In our system, we consider an initial state which is the direct product of a coherent spin state polarized along $-\hat{z}$ and a coherent state of the cavity field, $\ket{\psi_0}=\ket{(-N/2)_z}\otimes\ket{\alpha}$ where $\hat{S}_z\ket{(-N/2)_z} = (-N/2)\ket{(-N/2)_z}$ and $\hat a\ket{\alpha}=\alpha \ket{\alpha}$. Time evolution of this initial state 
under Eq.~(\ref{IIHamiltonian}) generates the coherent superposition
\begin{equation}\label{genericCat}
    \ket{\psi^{\mathrm{AL}}_{\mathrm{cat}}} = \sum_{m=-N/2}^{N/2} c_m \vert m_x \rangle \otimes \vert \alpha e^{-i \omega_m t} \rangle ,
\end{equation}
where $\omega_m = \chi m$, $c_m$ are the expansion coefficients of the state $\ket{(-N/2)_z}$ in the basis $\hat{S}_x\ket{m_x} = m_x\ket{m_x}$ and the superscript AL emphasizes that the state is a generalized cat-state of \emph{both} the atomic and light degrees of freedom. Generalized cat-states are widely known to be an excellent resource for quantum metrology due to  their fine structure in phase-space \cite{Zurek2001,Toscano_2006}, inversely proportional to the characteristic separation of the coherent amplitudes $\sim 1/\vert\alpha\vert$, which makes a perturbed state rapidly orthogonal to the initial cat. This is illustrated in Fig.~\ref{fig:CavityAndWigner}(b) where we plot the Wigner function of the equivalent generalized cat-state associated with just the bosonic degree of freedom, $\vert \phi_{\mathrm{cav}}\rangle \propto \sum_{m = -N/2}^{N/2} c_m \vert \alpha e^{-i\omega_m t} \rangle$. The Wigner function displays increasing detail as the superposed coherent states $\vert \alpha e^{-i \omega_m t} \rangle$ disperse in time \cite{PRA}.

\begin{figure}[!t]
    \includegraphics[width=8.6cm]{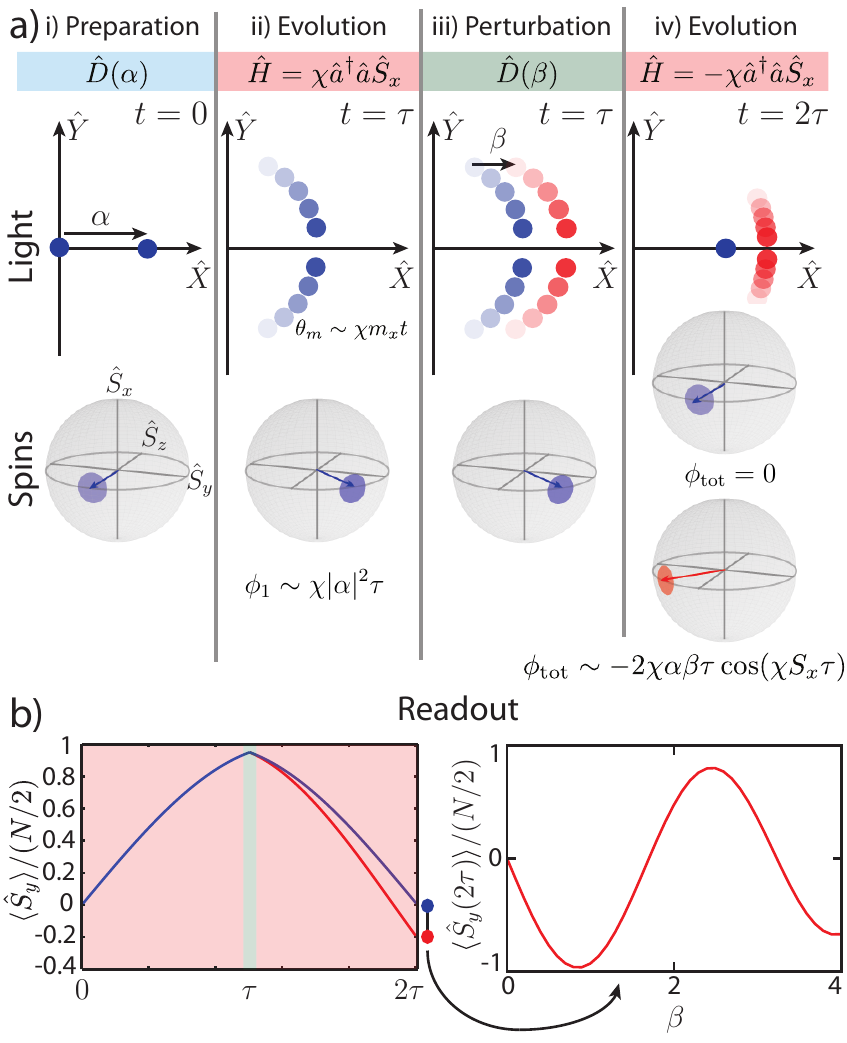}
    \caption{
    (a) Preparation of generalized cat-state $\vert\psi^{\mathrm{AL}}_{\mathrm{cat}}\rangle$ and interferometric protocol.
    (i) The cavity is injected with a coherent field $\alpha$ and the collective spin is fully polarized along $-\hat{z}$ (blue circles). (ii) Fluctuations in the spin projection combined with the dispersive interaction drive a rotation of the initial bosonic coherent state into a superposition at angles $\theta_m \sim \chi m_x \tau$. Conversely, the large cavity occupation rotates the collective Bloch vector by $\phi_1 = \sim \chi \vert\alpha\vert^2 \tau$ about $\hat{x}$. (iii) The cavity field is coherently displaced by $\beta$ (red circles). The spin degree of freedom is unaffected. (iv) By reversing the sign of the dispersive interaction the initial rotations are undone. If $\beta \neq 0$ the final cavity state (red circles) does not return to the original coherent state. Similarly, the collective spin rotates back under the evolution by $\phi_2 \sim -\chi\vert\alpha  e^{-i\chi S_x \tau} + \beta\vert \tau$ about $\hat{x}$, leading to an overall rotation $\phi_{\mathrm{tot}} = \phi_1 + \phi_2 \sim -2\chi\alpha\beta \tau\mathrm{cos}(\chi S_x \tau)$ relative to the initial state along $-\hat{z}$.
    (b) In the absence of a displacement the time-reversal revives the initial state (blue). However, perturbation of the cavity field destroys this revival, reflected in $\langle\hat{S}_y\rangle \neq 0$ for $\beta \neq 0$ (red). The dependence of the final $\langle\hat{S}_y\rangle$ on $\beta$ allows the parameter $\beta$ to be inferred.}
    \label{fig:ProtocolAndCavity}
\end{figure}

Quantitatively, the metrological utility of the state $\ket{\psi^{\mathrm{AL}}_{\mathrm{cat}}}$ to small displacements $\beta \equiv \vert \beta\vert e^{i\theta}$ can be characterized by the quantum Fisher information (QFI) $\mathcal{F}^{\theta}_Q = 4\langle (\Delta\hat{X}_{\theta})^2 \rangle$. Here, $\hat{X}_{\theta} = \hat{a}e^{-i\theta} + \hat{a}^{\dagger}e^{i\theta}$ is a bosonic quadrature operator which generates the displacement and $ \langle (\Delta \hat{X}_{\theta})^2 \rangle \equiv  \langle \hat{X}_{\theta}^2 \rangle - \langle \hat{X}_{\theta} \rangle^2 $ its variance. The QFI is related to the quantum Cramer-Rao bound \cite{Braunstein1994} $(\delta\beta)^2 \geq 1/\mathcal{F}^{\theta}_Q$, which describes the precision $\delta\beta$ to which a displacement can be estimated. For short times, $t\ll 1/(\vert\chi\vert\sqrt{N})$, the QFI is maximal for displacements parallel to the initial coherently displaced state $\alpha$, i.e. $\theta = 0$, and is given by $\mathcal{F}^0_Q \approx 4(1 + N\chi^2\vert\alpha\vert^2 t^2)$. For comparison, the SQL is defined as the sensitivity achievable with the original coherent state $\vert \alpha \rangle$, $(\delta\beta)^2 = 1/4$, equivalent to a QFI of $\mathcal{F}_{Q,{\vert \alpha \rangle}} = 4$. A key aspect of the metrological gain provided by a generalized cat-state is the characteristic growth rate of $\mathcal{F}^0_Q$, which is both collectively enhanced $\propto \sqrt{N}$ and increases with the coherent amplitude $\propto \vert\alpha\vert$. At longer times $\chi t\gtrsim 1/\sqrt{N}$, the atomic fluctuations will superpose the bosonic coherent state completely about a circle of radius $\vert\alpha\vert$ in phase-space [Fig.~\ref{fig:ProtocolAndCavity}(a)]. This state is sensitive to perturbations along any direction with $\mathcal{F}^{\theta}_Q \approx 4 + 8\vert \alpha \vert^2$.

\noindent{\textit{Protocol:}} While the QFI bounds the optimal sensitivity achievable with a given quantum state, in practice the sensitivity is limited by the measurements which can be implemented in the experimental platform \cite{Toth_2014}. Specifically, the phase-space structure of cat-states like $\vert \psi^{\mathrm{AL}}_{\mathrm{cat}}\rangle$ [see Fig.~\ref{fig:CavityAndWigner}(b)] which makes them highly sensitive to displacements also requires complex measurements necessitating single-particle resolution \cite{Bollinger_1996} or the ability to perform state tomography \cite{Toscano_2006,Zurek2001}. Predictably, then, measurements of simple cavity observables like the quadrature $\hat{X} = \hat{a} + \hat{a}^{\dagger}$ are not useful to sense the small perturbation of $\vert \psi^{\mathrm{AL}}_{\mathrm{cat}}\rangle$ \cite{PRA}. Recently, it has been demonstrated that a powerful solution is to use a time-reversal protocol \cite{Yurke_1986,Hudelist_2014,Linnemann_2016,Hosten_2016,Penasa_2016,Macri_2016,Davis_2016,Szigeti_2017,Fabian_2018,Huang_2018,Wrubel_2018,Burd2019}, wherein the initial entangling dynamics are reversed after the perturbation. If the initial prepared state is Gaussian, such as $\vert (-N/2)_z \rangle \otimes \vert \alpha \rangle$ used here, then even simple observables can typically be used to achieve almost ideal sensitivity to the perturbation. Moreover, reversal protocols can have other favourable properties including robustness to experimental detection noise \cite{Davis_2016,Nolan_2017,Haine_2018,Mirkhalaf_2018}. 

Our proposed interferometric protocol consists of (see Fig.~\ref{fig:ProtocolAndCavity}): (1) prepare the cavity in a coherent state of real amplitude $\alpha$ and with pseudospin pointing along $-\hat{z}$, (2) evolve with $\hat{H}$ for time $\tau$, (3) coherently displace the cavity by $\beta$, (4) evolve with $-\hat{H}$ for time $\tau$, and (5) measure an observable $\hat{M}$ (at final time $2\tau$). We will demonstrate that measuring spin observables, e.g., $\hat{M}=\hat{S}_y$, combined with this protocol are sufficient to nearly saturate the Cramer-Rao bound. For completeness, we have found that other measurements such as the occupation or quadratures of the cavity field are not sensitive observables \cite{PRA}.

Physical intuition for the protocol and the choice of $\hat{M}=\hat{S}_y$ can be gained from a semi-classical picture [see Fig.~\ref{fig:ProtocolAndCavity}]. The first evolution can be interpreted as a rotation of the collective spin about $\hat{x}$ by an angle $\phi_1 \sim \chi\vert\alpha\vert^2 \tau$, driven by the large coherent occupation of the cavity. After the perturbation of the cavity field, the reverse evolution counter-rotates the spin by $\phi_2 \sim -\chi\vert\alpha e^{-i\chi S_x \tau} + \beta\vert^2 \tau$, where the phase of the $\alpha$ term accounts for the evolution of the bosonic mode in the first stage and $S_x$ is a semi-classical fluctuation of characteristic scale $\sim \sqrt{N}$ due to quantum projection noise. Coarsely, this second rotation over-compensates for the first, leading to a small net rotation $\phi_{\mathrm{tot}} = \phi_1 + \phi_2 \sim -2\chi\alpha\beta \tau\mathrm{cos}(\chi S_x \tau)$. This rotation is measurable in collective spin observables and the $\beta$-dependence is amplified by the coherent amplitude $\alpha$. Finally, measuring $\hat{M} = \hat{S}_y$ we find an attainable sensitivity
\begin{equation}
    (\delta\beta)^2\approx\frac{1}{4N\chi^2 \tau^2\alpha^2} ,
\end{equation}
where we assume $\beta \to 0$ and $\tau \lesssim (\chi \sqrt{N})^{-1}$. The divergence at early times is a consequence of the spin projection noise $\propto \sqrt{N}$: The atoms and cavity must interact for sufficiently long so that the displacement $\beta$ induces a resolvable rotation in the collective spin greater than $1/\sqrt{N}$~rad (the spin SQL).

\noindent{\textit{Engineered atom-light interaction:}} The dispersive interaction of Eq.~(\ref{IIHamiltonian}) can be engineered via two approaches, both starting from the underlying Tavis-Cummings model which describes the uniform coupling of a single bosonic mode to a collection of $N$ two-level atoms:
\begin{equation}\label{IIITavis-Cummings}
    \hat{H}_{\mathrm{TC}}=g (\hat{a}^{\dagger}\hat{S}^-+\hat{a}\hat{S}^+)-\Delta_c \hat{a}^{\dagger}\hat{a} .
\end{equation}
Here, $2g$ is the single-photon Rabi frequency and $\Delta_c$ is the detuning of the cavity mode from the atomic transition. We highlight that this Hamiltonian could also be realized by driving the sideband transition in a trapped ion array with uniform coupling to a single motional mode \cite{Bohnet2016,Bollinger_2018}. 

The most obvious method to engineer the dispersive interaction is to follow the approach used in microwave cavity platforms and work in the limit of a large detuning, $|\Delta _c|\gg |g| \sqrt{N}$. However, the large detuning that would be required leads an unfavourable scaling of the coherent interaction scale $\chi \sqrt{N}$ relative to typical cavity loss rates \cite{PRA}.

Here, we instead outline a protocol to engineer a dispersive coupling by tuning the cavity to resonance $\Delta_c = 0$ \cite{Penasa_2016,PRA} and injecting a large coherent state $\braket{\hat{a}^{\dagger}\hat{a}}=|\alpha|^2\gg 1$. To elucidate that this leads to a dispersive interaction we first adopt a number-phase representation $\hat{a}=\sqrt{\hat{a}^{\dagger}\hat{a}}e^{i\hat{\phi}}$ \cite{Susskind1964}. In a manner analogous to a classical phase, the phase operator can be absorbed into the spin operators \cite{PRA}, which amounts then to a spin rotation about $\hat{z}$. The phase noise added to the spin quadratures ($\sim N|\alpha|^{-1}$ for a coherent state) is negligible compared to the initial spin projection noise ($\sim \sqrt{N}$) so it can be neglected. Hence the dynamics is initially driven by number fluctuations $\delta\hat{n}=\hat{a}^{\dagger}\hat{a}-|\alpha|^2$. Keeping terms of first order in $\delta\hat{n}$, we can perform the replacement $\hat{a} \to \sqrt{\vert\alpha\vert^2 + \delta\hat{n}} \to \vert\alpha\vert + \delta\hat{n}/(2\vert\alpha\vert)$. Manipulation of $\hat{H}_{\mathrm{TC}}$ under this approximation yields
\begin{equation}
    H_{\mathrm{R}}=g\vert\alpha\vert\hat{S}_x+\frac{g}{\vert\alpha\vert}\hat{S}_x\hat{a}^{\dagger}\hat{a} . \label{eqn:ResHam}
\end{equation}
The first term and mean field contribution $\propto |\alpha|^2$ of the second term describe Rabi-flopping of the atomic transition, which is intuitively expected when the cavity is driven with a strong resonant field. The dispersive coupling, equivalent to Eq.~(\ref{eqn:ResHam}) with $\chi = g/\vert\alpha\vert$, is a result of our more detailed treatment incorporating quantum fluctuations. We expect this expansion to be valid when $\delta \hat{n}\ll \braket{\hat{a}^{\dagger}\hat{a}}$. As a more concrete condition, as the atom-light interaction in $\hat{H}_{\mathrm{TC}}$ can in principle facilitate an exchange of up to $N$ excitations between the bosonic and spin degrees of freedom we require $N\ll\vert\alpha\vert^2$. For completeness, the sign of the Hamiltonian $\hat{H}_{\mathrm{R}}$ can be reversed by a global rotation about $\hat{z}$ so that $\hat{S}_x \to -\hat{S}_x$ and $\hat{S}_y \to -\hat{S}_y$ \cite{PRA}.


We note that a dispersive interaction of the form Eq.~(\ref{IIHamiltonian}) has previously been engineered in hyperfine states coupled via an optical cavity using Raman transitions \cite{Schleier-Smith2010b,Hosten_2016,Braverman_2019}. However, this approach is incompatible with our goal of generating non-classical states of light inside the cavity, as engineering the interaction required a coupling to a short-lived intermediate state that is driven by continuous pumping of the cavity by an external field.

\noindent{\textit{Effects of noise and dissipation:}}
Of crucial concern to any realistic quantum sensor is a complete characterization of sources of technical noise and decoherence, and their effects on the sensors performance. For our system, intrinsic sources of decoherence are photon loss from the cavity at rate $\kappa$ and single-particle spontaneous emission of the atoms at rate $\gamma$. The former mechanism is important in optical cavities, as experiments typically operate in the limit  $g\ll \kappa$. This is to be contrasted with microwave cavities which can operate in a strong coupling limit with a single qubit \cite{Schuster_2007,Girvin_2014}. In the following we will demonstrate that decoherence in an optical cavity can still be overcome by harnessing the collective enhancement of an $N$ atom ensemble.

Leakage of photons is primarily through the cavity mirrors at a rate $\kappa\sim0.1$-$1$~MHz in state-of-the art experiments \cite{Norcia_2018,Braverman_2019,Davis_2019,Vaidya_2018}. The impact of photon loss on the protocol can be estimated from a toy-model based on the archetypal bosonic cat-state $\vert\psi_{\mathrm{cat}}\rangle = (\ket{\alpha_0} + \ket{-\alpha_0})/\sqrt{2}$. Photon loss destroys the superposition (i.e. off-diagonal coherences) exponentially with the separation $\vert\alpha_0\vert$ of the coherences, $e^{-2\kappa |\alpha_0|^2t}$. This characteristic decay is similarly displayed by the QFI with respect to small displacements, $\mathcal{F}^{\mathrm{B}}_Q \approx 4+16\vert\alpha_0\vert^2e^{-\kappa t}e^{-4\kappa\vert\alpha_0\vert^2 t}$ \cite{PRA}.

Whilst the exponential decay of this toy model indicates the generalized atom-light cat-state is fragile, we find there is a relatively large region of parameter space in which the effects of dissipation, though negative, are not overtly detrimental to our protocol. Specifically, as entanglement and coherences are generated dynamically (via Hamiltonian evolution), simultaneously with photon loss, for an initially unentangled product state we expect that there is an optimum time at which $\mathcal{F}^0_Q$ is maximized. As a crude estimate we simplify Eq.~(\ref{genericCat}) by considering the relevant coherent state amplitudes to be those that are entangled with spin components $|m_x|<\sqrt{N}$, with a dynamically evolving cat separation $\alpha_0 \approx  \chi \sqrt{N}\alpha t$. Plugging this into the QFI prediction of $\vert\psi_{\mathrm{cat}}\rangle$, assuming $\kappa t\ll 1$ and minimizing over $t$ we obtain
\begin{equation}
    (\mathcal{F}^0_Q)_{\text{opt}}-4\sim \bigg(\frac{\chi^2\alpha^2N}{\kappa^2}\bigg)^{1/3},\hspace{0.3cm} t_{\text{opt}}\sim\bigg(\frac{1}{\kappa\chi^2N\alpha^2}\bigg)^{1/3}.
\end{equation}
A detailed analysis shows that this simple  estimation is correct, up to numerical prefactors \cite{PRA}. Particularly, even if $\chi \ll \kappa$, as is usually the case in optical systems, $\chi\vert\alpha\vert\sqrt{ N}>\kappa$ is enough to obtain a meaningful QFI. This scaling is also reflected in the sensitivity achievable with collective spin measurements. Specifically, for $\kappa \tau \ll 1$ the sensitivity is given as \cite{PRA}
\begin{equation}\label{eqn:kappasensitivity}
    (\delta\beta)^2_{\kappa} \approx \frac{1 + \frac{2}{3}\kappa (\sqrt{N}\chi\vert\alpha\vert \tau)^2 \tau}{4N\chi^2\vert\alpha\vert^2 \tau^2} = \frac{1}{4N\chi^2\vert\alpha\vert^2 \tau^2} + \frac{\kappa \tau}{6} ,
\end{equation}
which has the optimum
\begin{equation}\label{IVSensitivity}
     (\delta\beta)^2_{\kappa,\text{opt}}=\frac{1}{4}\bigg(\frac{3\kappa^2}{\chi^2N\alpha^2}\bigg)^{1/3},\hspace{0.3cm}\tau_{\text{opt}}=\bigg(\frac{3}{\kappa\chi^2N\alpha^2}\bigg)^{1/3}.
\end{equation}
This shows identical scaling to the Cramer-Rao bound $1/(\mathcal{F}^0_Q)_{\text{opt}}$. For our scheme $\chi = g/\vert\alpha\vert$, so the $\alpha$ factors cancel but $N$-fold enhancement remains, $(\delta\beta)^2_{\kappa,\mathrm{opt}} = (1/4)[3\kappa^2/(g^2 N)]^{1/3}$. This also grants some freedom to tune $\alpha$ to guarantee the validity of Eq.~(\ref{eqn:ResHam}).

\begin{figure}
    \includegraphics[width=8cm]{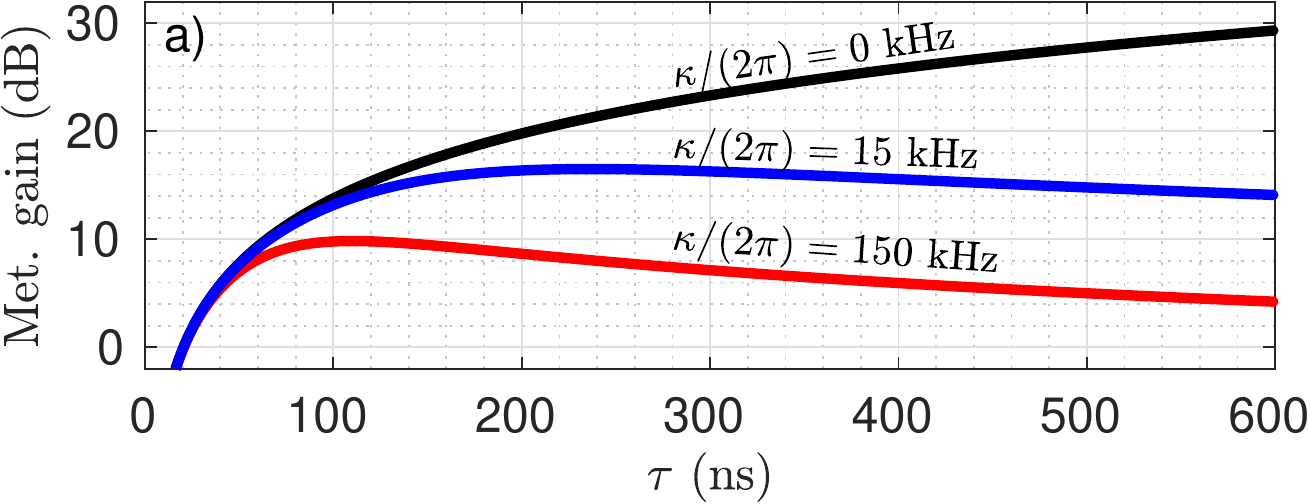}
    \includegraphics[width=8cm]{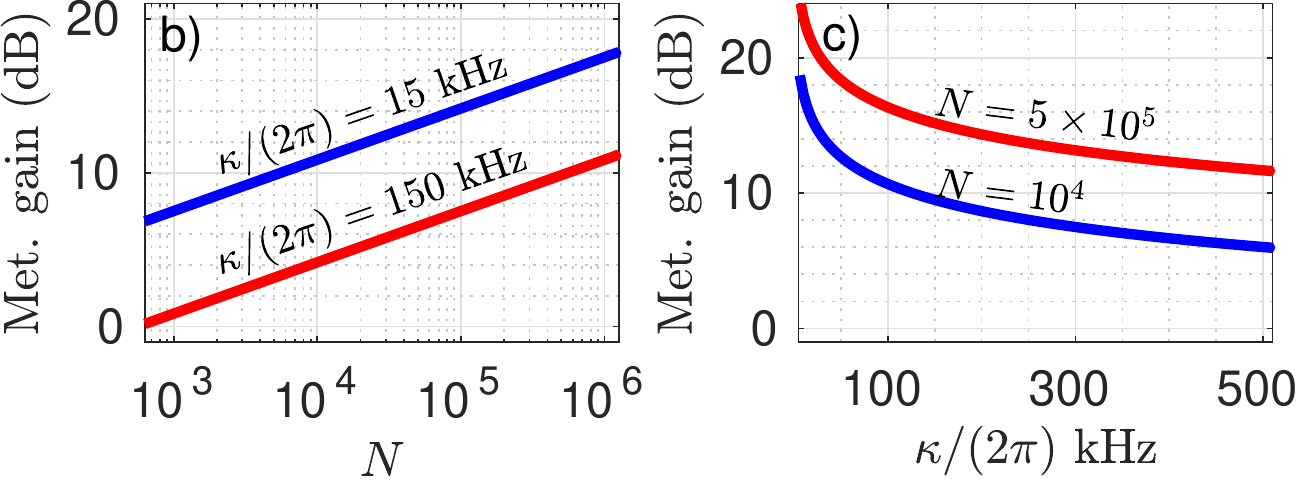}
    \caption{(a) Metrological gain $(\delta\beta)^2_{\mathrm{SQL}}/(\delta\beta)^2 \equiv [4(\delta\beta)^2]^{-1}$ as a function of interaction time $\tau$, for a range of cavity decay rates $\kappa$. The cavity is coupled resonantly, $N=5\times10^5$, $\alpha = 100\sqrt{N}$ and $g/(2\pi) = 11$~kHz for the $^1$S$_0 \to {}^3$P$_1$ transition of $^{88}$Sr. Optimal gain (maximised over $\tau$) is also plotted for: (b) $N$ and (c) $\kappa$. }
    \label{fig:SensitivityExamples}
\end{figure}

For current state-of-the-art experiments using long-lived optical transitions $\gamma \ll \kappa$, and spontaneous emission leads predominantly to a single-particle decay of spin observables on time-scales $1/\gamma$ \cite{PRA}. Whilst we present a full treatment in Ref.~\cite{PRA}, the relative dominance of photon decay [$\tau_{\mathrm{opt}} \ll 1/\gamma$ in Eq.~(\ref{IVSensitivity})] then means that we can safely neglect spontaneous emission in our quantitative examples. 



\noindent{\textit{Experimental realization:}} For concreteness, we present an example calculation for the optical cavity described in Refs.~\cite{Norcia2016,Norcia_2018}, where a single cavity mode is resonantly coupled to an ensemble of $N$ atoms trapped in a standing wave optical lattice oriented along the cavity axis. We assume uniform coupling of the atoms to the cavity mode, which can be realized via site-selective loading in the spatial lattice or using a ring cavity. Relevant experimental parameters are: $N\sim 10^5$-$10^6$ atoms, $\alpha = 100\sqrt{N}$ and $g/(2\pi)=11$~kHz for the $^1S_0\rightarrow{}^3P_1$ transition in $^{88}$Sr \cite{Norcia2016,Norcia_2018}. Preparation of the initial state spin state $\vert (-N/2)_z \rangle$ is via optical pumping to the $^{1}$S$_{0}$ ground-state, while the coherent state $\vert \alpha \rangle$ is injected via a laser. The spin projection $\hat{S}_y$ is mapped into atomic inversion ($\hat{S}_z$) by global rotations and measured by fluorescence \cite{Norcia_2018}. In Fig.~\ref{fig:SensitivityExamples}(a) we show the metrological gain over the SQL for $\kappa/(2\pi) = (0,15,150)$~kHz.

\noindent{\textit{Conclusion:}} We have demonstrated that atom-light interactions in an optical cavity can be utilized to generate highly non-classical states for quantum metrology in the optical domain. The collective atom-light interaction is pivotal for not only the generation of entanglement but also for readout. Whilst the examples presented in this work have focused on optical cavities, we stress that the methods analyzed here can be readily applied to  other spin-boson systems including trapped ions \cite{Bollinger_2018}, microwave cavities \cite{Delglise_2008}, circuit-QED \cite{Wallraff_2009,Viennot_2018,Fink2009} and other hybrid quantum systems \cite{Kolkowitz_2012,Aspelmeyer_2014}, with immediate applications to the sensing of weak forces \cite{Gilmore2017,Burd2019}.

\begin{acknowledgments}
\noindent{\textit{Acknowledgements:}} We acknowledge helpful discussions with J. Bollinger and B. Brubaker during the preparation of this manuscript. This work is supported by the AFOSR grant FA9550-18-1-0319 and its MURI Initiative, by the DARPA and ARO grant W911NF-16-1-0576, the ARO single investigator award W911NF-19-1-0210,  the NSF PHY1820885, NSF JILA-PFC PHY-1734006 grants, and by NIST. 
\end{acknowledgments}

\bibliography{library}

\end{document}